\documentclass[epj]{svjour}
\usepackage{latexsym}
\usepackage{graphicx}
\usepackage{mathptmx}      
\usepackage{flushend}
\usepackage[numbers,sort&compress]{natbib}
\usepackage[colorlinks,citecolor=blue,urlcolor=blue,linkcolor=blue]{hyperref}
\usepackage[utf8]{inputenc}

\usepackage[italic]{hepnames}

\usepackage{breqn}             
\makeatletter                  
\let\cat@comma@active\@empty   
\makeatother                   

\usepackage{graphics}
\begin{document}
\title{$B_c$ meson spectroscopy motivated by general features of pNRQCD}
\author{Raghav Chaturvedi$^{1}$ \and Ajay Kumar Rai$^{2}$
}                     
\offprints{raghavr.chaturvedi@gmail.com}          
\institute{$^{1}$Ministry of Education, Dubai, \textit{U.A.E}\\
$^{2}$Department of Applied Physics, Sardar Vallabhbhai National
Institute of Technology, Surat, Gujarat, {\it INDIA}}
\date{Received: date / Revised version: date}

\abstract{In the present article the mass spectrum, decay constant, weak decay widths, life time and branching fraction ratios and electromagnetic transition widths are calculated for ground and radially excited $B_c$ meson. To calculate the above properties the Schr\"{o}dinger equation has been solved numerically for the potential. The potential employed consists of relativistic correction in the framework of pNRQCD, added to the Cornell potential. The calculated results are compared with available experimental and theoretical results.}

\maketitle

\section{Introduction}
\label{sec:Indroduction}
Lately, many new resonances have been discovered in the hadronic family, which also include the excited heavy $B_c$ meson. The ground state of the $B_c$ meson was first observed in 1998 at Fermi lab by CDF collaboration\cite{CDF:1998ihx} with mass $6.40 \pm 0.39 \pm 0.13$            GeV, the observation was later supported by Abulencia $et. al.$\cite{CDF:2005yjh}, Aaltonen $et. al.$\cite{CDF:2007umr}, D0 Collaboration \cite{D0:2008bqs}, and later LHCb\cite{LHCb:2012ihf}.
After a dull of few years, in 2014 a structure was reported by ATLAS collaboration with mass of $6842 \pm 9$ MeV\cite{ATLAS:2014lga}, Which was consistent with the value predicted for $B_c (2S)$.
Recently, CMS and LHCb collaboration have observed $B_c (2^1 S_0)$ and $B_c (2^3 S_1)$ states in $B_c^+ \pi^+ \pi^-$ invariant mass spectrum, their determined masses are $6872.1 \pm 2.2$ and $6841.2 \pm 1.5$ MeV, respectively\cite{ATLAS:2014lga,CMS:2019uhm,LHCb:2019bem,CMS:2020rcj}. These latest developments in the discovery of radially excited states of $B_c$ meson, coupled with the fact that the experimental information about the $B_c$ meson is still insufficient, has sparked more interest both within theoretical\cite{Martin-Gonzalez:2022qwd,Wang:2022cxy,Tang:2022xtx,Mansour:2021rru} and experimental groups, with more research on $B_c$ meson expected to be reported in the near future.

Because of its unique composition of two different heavy quarks, leading to its high stability, the $B_c$ meson enjoys lots of theoretical attention related to its production\cite{Chang:1992bb,Chang:1991bp,Braaten:1993jn,Cheung:1993qi,Chang:1992jb}, decays\cite{Chang:2001pm,Colangelo:1999zn,Qiao:2012hp,Ivanov:2006ni,Chang:1992pt,Kiselev:1993ea,Liu:1997hr, Kiselev:2000pp}, and mass spectrum\cite{Gregory:2009hq,Dowdall:2012ab,Godfrey:1985xj,Chen:1992fq,Fulcher:1998ka,Gershtein:1994dxw}.

Many theoretical approaches like the relativistic quark models \cite{Godfrey:1985xj,PhysRevD.52.5229,Gupta:1995ps,Ebert:2002pp,Ikhdair:2003ry,Godfrey:2004ya,Ikhdair:2004hg}, qcd sum rules\cite{ATLAS:2021moa,Wang:2013cha}, continuum functional methods for QCD\cite{Chang_2021,Chen:2020ecu,Yin:2019bxe}, effective field theories\cite{Brambilla:2000db,Penin:2004xi,Peset_2018,Peset:2018jkf}, lattice QCD\cite{Allison:2004be,Mathur:2018epb,Dowdall:2012ab}, relativistic and non-relativistic phenomenological models have tried to explain the production and decays of quarkonia and $B_c$ mesons.

Because of the presence of two heavy quarks in $B_c$ meson, non-relativistic phenomenological potential models remain best option for their analysis. In non-relativistic potential models,  exists several forms of quark anti-quark potential in literature, like Bethe-Salpeter approach\cite{Sauli:2011aa,Leitao:2014jha,Fischer:2014cfa}, screened potential model \cite{Deng:2016stx,Deng:2016ktl}, constituent quark model\cite{Segovia:2016xqb}, linear plus confinement non-relativistic potential model \cite{Soni:2017wvy}, non-linear potential model \cite{Devlani:2014nda}.

It is essential for any potential model to compute the mass Spectra and the decay properties of $B_c$ meson using parameters fitted for heavy quarkonia. Considering from the first principles of QCD it is difficult to derive the interaction potential for full range of quark anti-quark separation. In this article we have employed one gluon exchange plus linear confinement, this form of potential is also supported by lattice QCD, to this potential we add relativistic correction term which has been derived in the framework of pNRQCD\cite{Brambilla:2010cs,Brambilla:2014jmp}, then the Schr\"{o}dinger equation has been solved numerically using the Mathematica notebook. pNRQCD  is an EFT which is generally classified in the powers of the inverse of the heavy quark mass or velocity. In quarkonia, prevails a pecking order $m\gg mv \gg mv^2$, with $m \gg \Lambda_{QCD}$. To have better command of such pecking order the EFT’s are employed for high accuracy calculations. Mainly two EFT’s, NRQCD and pNRQCD have been used in the past, NRQCD is derived by integrating the energy scale above $"m"$ in QCD, and pNRQCD but integrating further the energy scale above $"mv"$ in NRQCD. A higher energy contribution is incorporated in ineffective couplings, called the matching coefficient.
Using the same approach as mentioned here, we have calculated the mass Spectra and decays of charmonium \cite{Chaturvedi:2019usm} and bottomonium \cite{Chaturvedi:2018tjr}. The quark masses and confining strengths were determined after fitting the spin averaged ground state masses with experimental data of respective mesons. Using the same quark masses and confining strength the mass spectra of $B_c$ meson has been calculated, and the decay properties have also been calculated using the determined wavefunctions and overlap integral. To validate the model, along with mass spectra calculations certain decay channels needs to be tested as well. The decay constant is an important parameter for week decays, in the present article it has been calculated with and without QCD correction. The total decay width and lifetime have also been calculated. To understand the spin-spin interaction the calculation of radiative decay width is also important, also the magnetic dipole transitions can help in identification of new excited $B_c$ meson states. Hence, the E1 and M1  transition widths have also been evaluated. In future at various experimental facilities, due to ever expanding bandwidth of collision energy and high luminosity many new states can be observed and detailed analysis of the observed states can be made possible.\\
After the introduction in section\ref{sec:Indroduction}, in section\ref{sec:Theoretical framework} we discuss theoretical framework for calculating the mass spectra and various decay properties of $B_c$ meson, section\ref{sec:Result} contains results and discussion, finally in section\ref{sec:Conclusion} we conclude the article.

\section{Theoretical framework}
\label{sec:Theoretical framework}

\subsection{Mass Spectroscopy}

The bound state of the two quarks can be studied both relativistically and non-relativistically, Bethe Sallpeter equation considering relativistic quantum field is solved only in the ladder approximations. Also, In harmonic confinement the Bethe Sallpeter equation is successful in low flavor sectors only \cite{Isgur:1978xj,VijayaKumar:1997pg}. Considering the fact, that in heavy quark bound state $m_{Q,\bar{Q}} \gg \Lambda_{QCD} \sim	|\vec{p}|$, $i.e.$ the momenta of the quark is significantly lower than the mass of the quark anti-quark system, the non-relativistic treatment is best option for analyzing the heavy quark bound state.
Here, for the study of $B_c$ meson the non-relativistic Hamiltonian \cite{Gupta:1994mw} is given in Eq.\ref{eq:Hamiltonian}, the same Hamiltonian has been used in calculating the mass spectra of charmonia\cite{Chaturvedi:2019usm} and bottomonia\cite{Chaturvedi:2018tjr}.

\begin{equation} \label{eq:Hamiltonian}
  H=M+\frac{P^2}{2 M_{cm}}+ V_{1}(r)+ V_{SD}(r)
  \end{equation}
Here, $M = m_Q + m_{\bar{Q}}$, and $M_{cm} = \frac{m_Q m_{\bar{Q}}} {m_{Q} + m_{\bar{Q}}}$ are the mass and reduced mass of the system, and \textbf{$\vec{P}$ is the momentum of the meson considered in center of mass frame, where $\vec{P_Q} = - \vec{P_{{\bar{Q}}}} = \vec{P} $ and can be calculated as per \cite{Soni:2017wvy}, in quantum mechanics the momentum is replaced by momentum operator}. The potential $V_{1}(r)$ term contains three contributions. The first term, $V_v(r)$ which includes $1/r$, is analogous to the Coulomb type interaction corresponding to the potential
induced between quark and anti-quark through one gluon exchange that dominates at small
distances. The second term, $V_s(r)$ is the confinement part in the potential with $A$ the confinement term as a model parameter, it becomes dominant at larger distances. The third term, $V_p(r)$ is the relativistic correction in the frame of pNRQCD\cite{article}.
\begin{eqnarray}
\label{eq:potential}
V_{1}(r) & = & V_v(r) + V_s(r) + V_p(r) \nonumber \\
 &=& -\frac{4\alpha_{s}}{3r}+ A r + \left( \frac{1}{m_b} + \frac{1}{m_c} \right) V^{(1)}(r)\\
V^{(1)}(r)&=& -\frac{9\alpha_{c}^2}{8r^2}+ a \log r
\end{eqnarray}
$\alpha_c$ and $\alpha_s$ are effective running coupling constant and strong running coupling constant respectively, $m_{b}$ and $m_{c}$ are masses of bottom and charm quark taken from our previous studies, $a$ is a potential parameter, not to be confused with $A$ which is confinement term, the value of confinement term$(A)$ for present work is average value of '$A$' used for calculating the mass spectra for charmonia and bottomonia. $\alpha_s$  can be calculated as \begin{eqnarray}\label{eq:running_coupling}
    \alpha_s (\mu^2) = \frac{4 \pi}{(11-\frac{2}{3} n_f) \ln (\mu^2/\Lambda^2)}
\end{eqnarray}
where $n_f$ is the number of flavors, $\mu$ is renormalization scale related to the constituent quark masses as $\mu = 2 m_Q m_{\bar Q}/(m_Q + m_{\bar Q})$ and $\Lambda$ is a QCD scale which is taken as 0.15 GeV by fixing $\alpha_s$ = 0.1185 \cite{Patrignani:2016xqp} at the $Z$-boson mass.
To compute the masses of radially and orbitally excited states, the confinement strength with charm and bottom quark masses are fine tuned to reproduce the experimental spin averaged ground state masses of charmonia and bottomonia, the averaged value of confinement strength and bottom and charm quark masses are then used to calculate the spin averaged ground state mass for $B_c$ meson. In literature, exists few works\cite{Parmar:2010ii,Rai:2008sc,Patel:2008na} which have used different values of confinement strengths for different potential indices.
We use the parameters listed in Table.\ref{potential parameters} to calculate the spin average mass of $B_c$ meson. To calculate masses of different $n^mL_J$ states according to different $J^{PC}$ values we add spin dependent part of one gluon exchange potential perturbatively.

\begin{eqnarray}
V_{SD}(r) & = &\frac{1}{\left({\frac{m_b + m_c }{2}}\right)^2} \bigg[V_{SS}(r)+ V_{L\cdot S}(r)+ \cr &&
V_T(r) \bigg [ S(S+1)-3(S\cdot \hat{r})(S\cdot \hat{r}) \bigg]\bigg]
\end{eqnarray}
Where the spin-spin, spin-orbital and tensor interactions are given as\cite{article}.
\begin{eqnarray}
V_{SS}(r) = \frac{8}{9 }\frac{\alpha_s }{m_b m_{\bar{c}}}\overrightarrow{S}_b \overrightarrow{S}_{\bar{c}} 4 \pi \delta^3(\overrightarrow{r}),
\end{eqnarray}

\begin{eqnarray}
V_{L\cdot S}(r) &=& \frac{C_s}{2 r} \frac{d}{dr} (V_v(r) + V_s(r)) + \cr && \frac{C_f}{r}\left[-\left(1-\epsilon\right)\sigma +\left(\frac{\alpha_c}{r^2}+ \epsilon \sigma\right)\right]
\end{eqnarray}

\begin{eqnarray}
V_{T}(r) &=& \frac{{C_f}^2}{3} \frac{3 a }{r^3}
\end{eqnarray}
Here, $C_f=\frac{4}{3}$, $\epsilon=-0.21$, \textbf{$a=0.26$} and $\sigma=3.8$. We take the quark masses and confinement parameter(A) which were fixed in our previous studies \cite{Chaturvedi:2019usm,Chaturvedi:2018tjr}, but the \textbf{potential parameters $\epsilon$, $a$ and $\sigma$} have been set in the present work to fix the masses for lowest lying state of $B_c$ meson.
The Schr\"{o}dinger equation has been solved numerically using Mathematica notebook utilizing the Runge-Kutta method. The computed mass spectra for $B_c$ meson can be found in Table.[\ref{Table:mass s,p},\ref{Table:mass d,f}]. Moreover, the masses for different states of $B_c$ meson have been computed with and without considering the relativistic correction from pNRQCD.\\

\begin{table*}
 \caption{Potential parameters for $B_c$ meson.}
 \begin{center}
 \begin{tabular}{cccc}
 \hline
  $\mathit{m_b}$ & $\mathit{m_c}$ & $A( c\bar{c})$ & $A(b\bar{b})$\\
 \hline
 4.81GeV & 1.321GeV  & 0.191 $\frac{\mathit{GeV}}{\mathit{fm}}$ & 0.211 $\frac{\mathit{GeV}}{\mathit{fm}}$\\
 \hline
 \end{tabular} \end{center}
 \label{potential parameters}
 \end{table*}

\subsection{Decay Properties}
\subsubsection{Leptonic Decay Constants}
To understand the weak decays in quarkonia and $B_c$ mesons leptonic decay constant play an important role.
In non-relativistic limit the decay constants are give by the Van Royen-Weiskopf formula\cite{VanRoyen:1967nq}.
\begin{eqnarray}
  {f}^2_{p/v} &=& 12 \frac{|{R_{nsP/V}(0)}|^2}{M_{nsP/V}} \bar{C}^2(\alpha_s) \\
  \bar{C}(\alpha_s) &=& 1 - \frac{\alpha_s}{\pi} \left( \delta^{P,V} - \frac{m_Q-m_{\bar Q}}{m_Q + m_{\bar Q}} ln \frac{m_Q}{m_{\bar Q}}\right)\\
  fcorr &=& {f}^2_{p/v}\bar{C}(\alpha_s)
\end{eqnarray}
$\bar{C}(\alpha_s)$ is the QCD correction factor\cite{Braaten:1995ej,Berezhnoy:1996an}.
Here, $R_{nsP/V}(0)$ denotes the numeric value of normalized reduced wave-function at origin for corresponding states, and $M_{nsP/V}$ represents the masses. $m_Q$ and $m_{\bar{Q}}$ are the masses for bottom and charm quarks. $\delta^P$ = 2 and $\delta^V$ = 8/3. Using the above equations we determine the leptonic decay constant values $f_{corr}$ for $B_c$ meson.

\subsubsection{Weak Decays}
Inclusion of two different quark flavors makes $B_c$ meson differ from charminia and bottomonia. Because of flavor symmetry, the ground state pseudoscalar meson cannot decay via strong or electromagnetic decays. The $B_c$ meson decays only through weak decays, hence have relatively longer lifetimes. As per the spectator model\cite{AbdEl-Hady:1998uiq}, the total decay width of the $B_c$ meson is mainly segregated into three classes. (i) Decay of $b$ quark considering $c$ quark as a spectator, (ii) Decay of $c$ quark considering $b$ quark as a spectator, and (iii) Annihilation channel. The total is width is given as,
\begin{equation}\label{eq:weak_Bc}
\Gamma (B_c \rightarrow X) = \Gamma (b \rightarrow X) + \Gamma(c \rightarrow X) + \Gamma (Anni)
\end{equation}
While calculating total decay width we have not considered intervention among them. The inclusive decay width of $b$ and $c$ quark is given by,
\begin{equation}
\Gamma (b \rightarrow X) = \frac{9 G_F^2 |V_{cb}|^2 m_b^5}{192 \pi^3}
\end{equation}
\begin{equation}
\Gamma (c \rightarrow X) = \frac{9 G_F^2 |V_{cs}|^2 m_c^5}{192 \pi^3}
\end{equation}
\begin{equation}
\Gamma (Anni) = \frac{G_F^2}{8 \pi} |V_{bc}|^2 f_{corr}^2 M_{B_c}  \sum \left( m_i^2 \left(1-\frac{m_i^2}{M_{B_c^2}}\right)^2 \right)C_i
\end{equation}
Here, $C_i= 3 |V_{cs}|^2$ for decay through $\bar{c}s$ channel.  $V_{cs}$ and  $V_{cb}$ are the CKM matrices and their values have been taken from the PDG, $G_F$ is the Fermi coupling constant, $f_{corr}$ and $M_{B_c}$ are the computed decay constant and mass for pseudoscalar state. The values of $m_c$ and $m_{\tau}$ is taken from PDG. \\
The life time($\tau$) of $B_c$ meson is found by the following relation $\tau = \frac{\hbar}{\Gamma}$, where $\Gamma$ is total width given as per Equation.\ref{eq:weak_Bc}, and value of $\hbar = 6.582 \times 10^{-25} GeV.sec$.\\
To find the branching ratio, we take the ratio of $\Gamma (b \rightarrow X)$, $\Gamma (c \rightarrow X)$, and $\Gamma (Anni)$ respectively with the total decay width $\Gamma (B_c \rightarrow X)$.

\subsubsection{Pure and Radiative Leptonic Decay Widths}
In principle the pure leptonic decay ($B_{c}{\longrightarrow} {\ell}{{\nu}_{\ell}}$) can help in determining the decay constant ($fcorr$) if the fundamental Cabibbo-Kobayashi-Maskawa matrix element($|V_{bc}|$) of the standard model is known. As the pure leptonic decay is helicity suppressed not enough events can be collected, so the determination of decay constant($fcorr$) becomes difficult. Contrariwise, the computed value of decay constant drom any model the pure leptonic decay width $\Gamma(B_{c}{\longrightarrow}{\ell}{\overline{\nu}_{\ell}})$ can be determined.\\
The pure leptonic decay width of pseudoscalar meson are helicity suppressed by factor of $m_{\ell}^{2}/m_{B_{c}}^{2}$, and it can be computed as per\cite{Chang:1997re}, \begin{equation}
\Gamma(B_{c}{\longrightarrow}{\ell}{\overline{\nu}_{\ell}})=
{\frac{G_{F}^{2}}{8\pi}}|V_{bc}|^{2}f_{corr}^{2}m_{B_{c}}^{3}{\frac
{{m_{\ell}}^{2}}{{m_{B_{c}}}^{2}}}
\left(1-{\frac{{m_{\ell}}^{2}}{{m_{B_{c}}}^{2}}}\right)^{2},
\end{equation}
Additionally, there's an extra photon emitted in leptonic decays making the radiative pure leptonic decays escape helicity suppression. Therefore, the radiative decay may
be comparable, even larger than the corresponding pure leptonic decays. To find the Radiative Leptonic Decay Widths we use,
\begin{eqnarray}
\Gamma(B_{c}{\longrightarrow}\gamma{\ell^-}{\overline{\nu}_{\ell}})&=&
{\frac{\alpha G_{F}^{2}|V_{bc}|^{2}}{2592 \pi^2}}
f_{corr}^{2}m_{B_{c}}^{3} \times
\cr &&
\left[ \left(3- \frac{m_{B_{c}}}{m_b}  \right)^2 + \left(3- \frac{m_{B_{c}}}{m_c}  \right)^2 \right]
\end{eqnarray}

\subsection{Electromagnetic Transitions}
\label{sec:EM}
Using the computed wave function we calculate the electromagnetic transition width, this will put to test the different parameters used to calculate the mass spectra, helping us to reach to a consensus regarding the effectiveness of the potential used in the present work. Electromagnetic transitions are broadly classified as electric and magnetic multipole expansions, their calculations can help to understand the non-perturbative regime of QCD. We calculate the E1 and M1 transition widths in the framework of pNRQCD. The E1 transitions follow  $\Delta L = \pm 1$ and $\Delta S = 0$  selection rules, while the M1 transitions follow $\Delta L = 0$ and $\Delta S = \pm 1$ selection rules. In non-relativistic limit, the radiative $E1$ and $M1$ transition widths are given by \cite{Brambilla:2010cs,Radford:2009qi,Eichten:1974af,Eichten:1978tg,Pandya:2014qma}.

\begin{eqnarray}
\Gamma(n^{2S+1}L_{iJ_i} \to n^{2S+1}L_{fJ_f} + \gamma) &=&
\frac{4 \alpha_e \langle e_Q\rangle ^2\omega^3}{3} (2 J_f + 1) \cr &&
S_{if}^{E1} |M_{if}^{E1}|^2
 \end{eqnarray}
\begin{eqnarray}
\Gamma(n^3S_1 \to {n'}^{1}S_0+ \gamma) = \frac{\alpha_e \mu^2 \omega^3}{3} (2 J_f + 1) S_{if}^{M1} |M_{if}^{M1}|^2
\end{eqnarray}
where, mean charge content $\langle e_Q \rangle$ of the $Q\bar{Q}$ system, magnetic dipole moment $\mu$ and photon energy $\omega$ are given by
\begin{equation}
\langle e_Q \rangle = \left |\frac{m_{\bar{Q}} e_Q - e_{\bar{Q}} m_Q}{m_Q + m_{\bar{Q}}}\right |
\end{equation}
\begin{equation}
\mu = \frac{e_Q}{m_Q} - \frac{e_{\bar{Q}}}{m_{\bar{Q}}}
\end{equation}
and
\begin{equation}
\omega = \frac{M_i^2 - M_f^2}{2 M_i}
\end{equation}
respectively. Also, the symmetric statistical factors are given by
\begin{equation}
S_{if}^{E1} = {\rm max}(L_i, L_f)
\left\{ \begin{array}{ccc} J_i & 1 & J_f \\ L_f & S & L_i \end{array} \right\}^2\\
\end{equation}
and
\begin{equation}
S_{if}^{M1} = 6 (2 S_i + 1) (2 S_f + 1)
\left\{ \begin{array}{ccc} J_i & 1 & J_f \\ S_f & \ell & S_i \end{array} \right\}^2 \left\{ \begin{array}{ccc} 1 & \frac{1}{2} & \frac{1}{2} \\ \frac{1}{2} & S_f & S_i \end{array} \right\}^2.
\end{equation}
The matrix element $|M_{if}|$ for $E1$ and $M1$ transitions can be written as
\begin{equation}
\left |M_{if}^{E1}\right | = \frac{3}{\omega} \left\langle f \left | \frac{\omega r}{2} j_0 \left(\frac{\omega r}{2}\right) - j_1 \left(\frac{\omega r}{2}\right) \right | i \right\rangle
\end{equation}
and
\begin{equation}
\left |M_{if}^{M1}\right | = \left\langle f\left | j_0 \left(\frac{\omega r}{2}\right) \right | i \right\rangle
\end{equation}

\section{Result and Discussion}
\label{sec:Result}
\subsection{Mass Spectroscopy}
Using the determined confinement strength, quark masses, and taking into consideration the Cornell potential, along with the relativistic correction in pNRQCD framework we calculate 1S-6S, 1P-4P, 1D-4D, 1F-2F masses of the $B_c$ meson. The Mass spectroscopy be found in Table.\ref{Table:mass s,p}$\&$\ref{Table:mass d,f}. We have used quark masses $(m_b \& m_c) $ from our previous work\cite{Chaturvedi:2018tjr}$\&$ \cite{Chaturvedi:2019usm}, and the confinement strength(A) for $B_c$ meson is the arithmetic mean of the confinements strengths for charmonium and bottomonium from the previous work. \\
A  statistical test called goodness of fit ($\chi^2/\textit{d.o.f}$ ) gives an idea about the accuracy of computed data.  As per literature it must be calculated for the potential parameters. We calculated $\chi^2/\textit{d.o.f}$ for the potential parameters while calculating charmonium and bottomonium masses and the calculated values were 1.553 and 0.764 respectively. The lower is the $\chi^2/\textit{d.o.f}$ value, the better are the chosen parameters. We believe that for the present article there is no need to calculate $\chi^2/\textit{d.o.f}$ value, because the quark mass and potential parameter are the arithmetic mean of the values for charmonium and bottomonium mass calculations.\\
In Table.\ref{Table:mass s,p}$\&$\ref{Table:mass d,f}, the first column contains calculated masses when only the Cornell potential is considered, the second column contains masses when relativistic correction to the Cornell potential in the framework of pNRQCD has been considered. It can be observed that the relativistic correction suppresses the masses in comparison to the masses calculated considering only the Cornell potential. The calculated masses from the present approach are in agreement with available experimental masses, and masses from available theoretical approaches like lattice QCD\cite{Davies:1996gi}, relativistic QM(RQM)\cite{Godfrey:2004ya,Ebert:2011jc}, non-relativistic potential models(NR)\cite{Devlani:2014nda,Soni:2017wvy}, screening potential\cite{Tang:2022xtx}, Salpeter approach\cite{Wang:2022cxy}. \\
The latest evolution is the finding of the first radially excited pseudoscalar $B_c(2S)$ and vector $B_c^{*}(2S)$ states. Theories and experiments predict an interesting phenomenon, $M(B_c^{*})- M(B_c)$ is larger than $M(B_c^{*}(2S))- M(B_c(2S))$. We calculate $M(B_c^{*})- M(B_c) = 70 MeV$ and $M(B_c^{*}(2S))- M(B_c(2S))= 23 MeV$, the experimental $M(B_c^{*}(2S))- M(B_c(2S))= 31 MeV$. For $M(B_c^{*}(1S))$, $M(B_c^{*}(2S))$, and $M(B_c(2S))$ states the masses calculated by us differ only by $3 MeV$, $15 MeV$, and $14 MeV$, respectively. The calculated masses for other states which have not been observed experimentally differ from the masses of theoretical approaches by $0.5 \%$ to $0.8\%$\\
After analyzing masses of radially and orbitally excited states. We construct $(n_r,M^2)$ and $(J,M^2)$ Regge trajectories. It is a tool to help assign higher radially and orbitally excited states to the $B_c$ meson family. We use the following definition, \begin{eqnarray}
 J = \alpha M^2 + \alpha_0\\
 n_r = \beta M^2 + \beta_0
\end{eqnarray}
$\alpha, \beta$ are the slopes and $\alpha_0, \beta_0$ are the intercepts.

\begin{figure}
\includegraphics[width=0.5\textwidth]{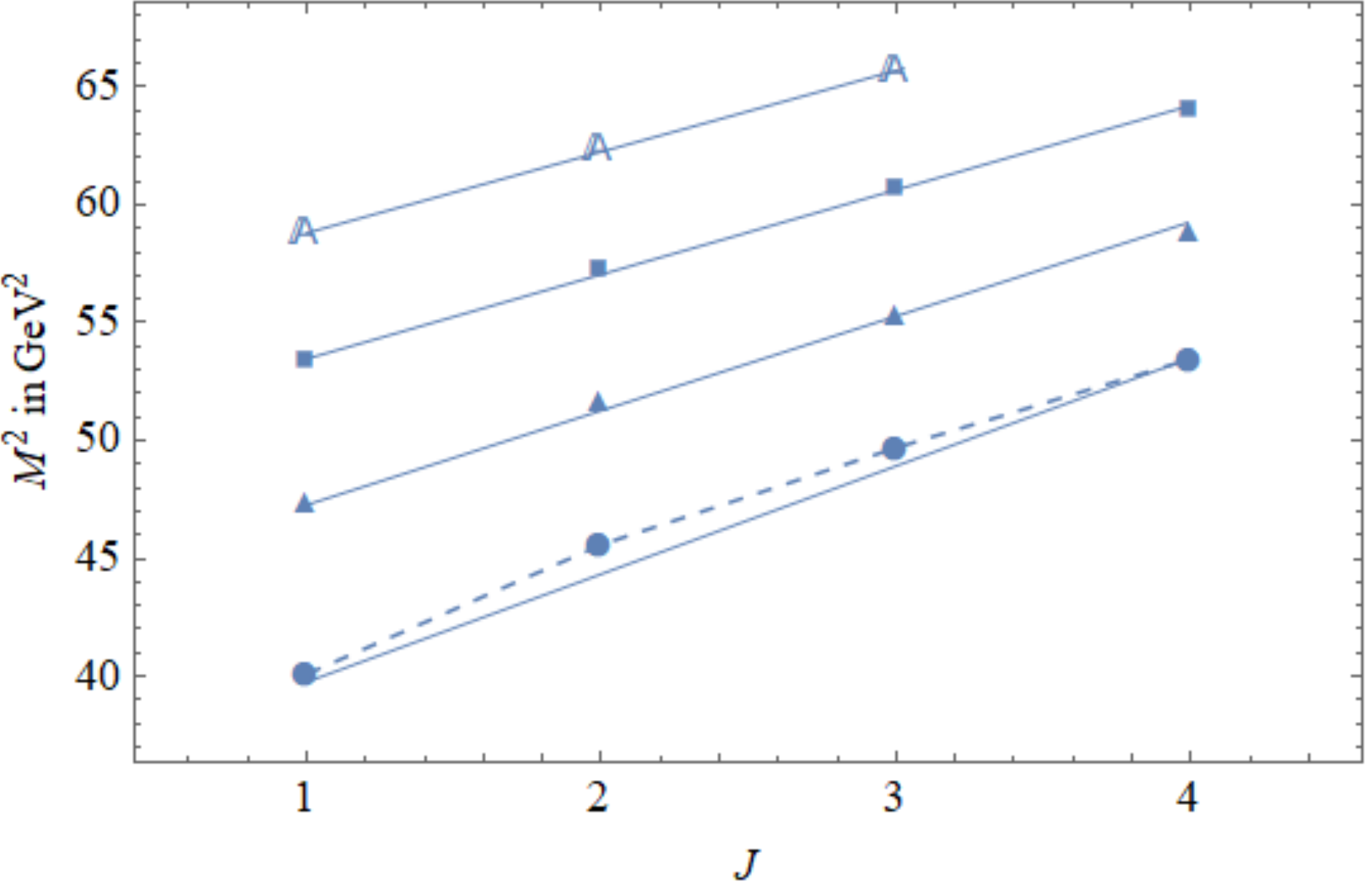}
\includegraphics[width=0.5\textwidth]{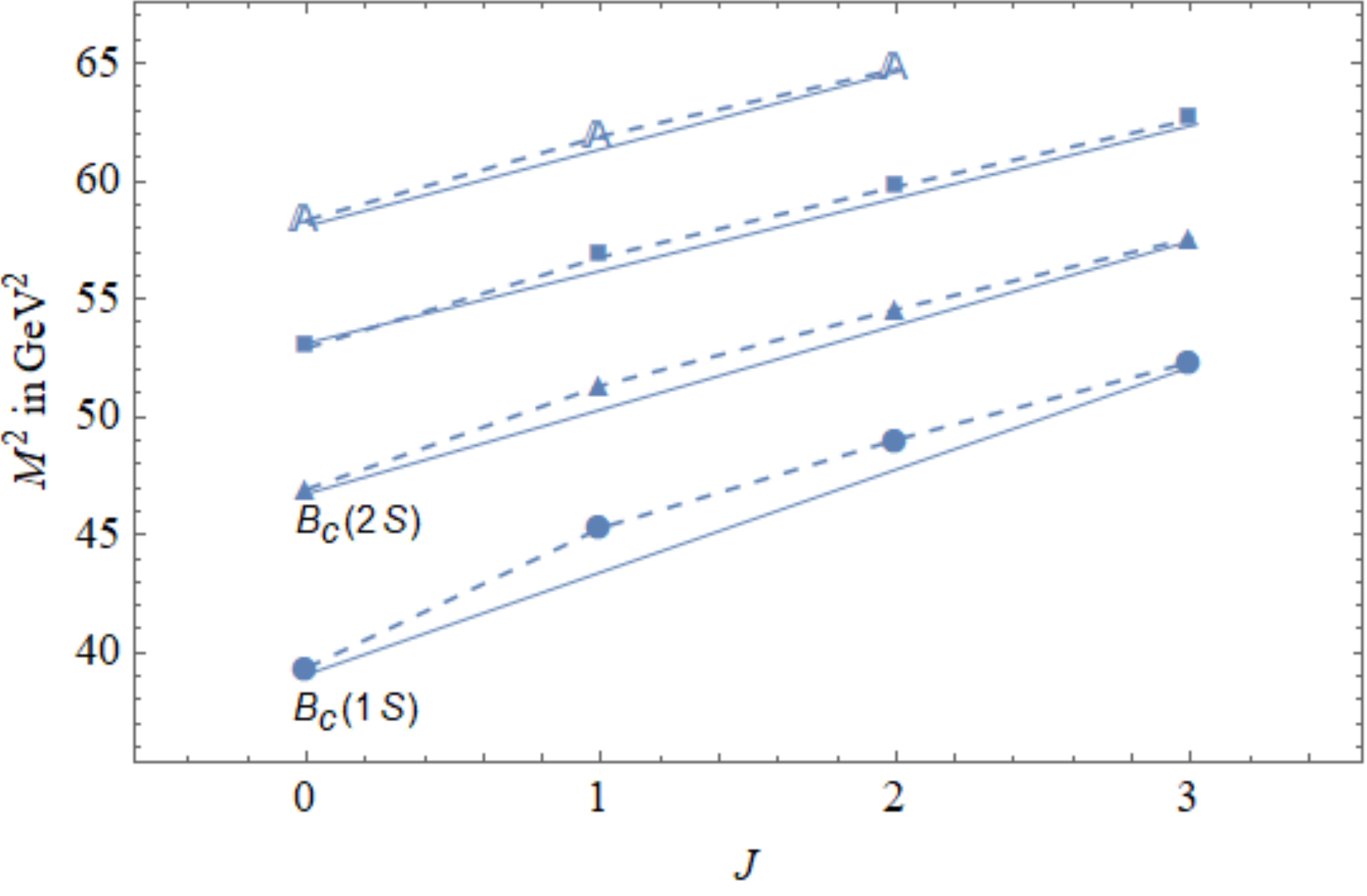}
\caption{$(J,M^2)$ Regge trajectory of parent and daughter for $B_c$ meson with natural parity($P = (-1)^J$) and unnatural parity($P = (-1)^{J+1}$). Solid dots indicates predicted mass, hollow shapes represent experimental masses.}
\label{fig:7}
\end{figure}

\begin{figure}
\centering
\includegraphics[width=0.5\textwidth]{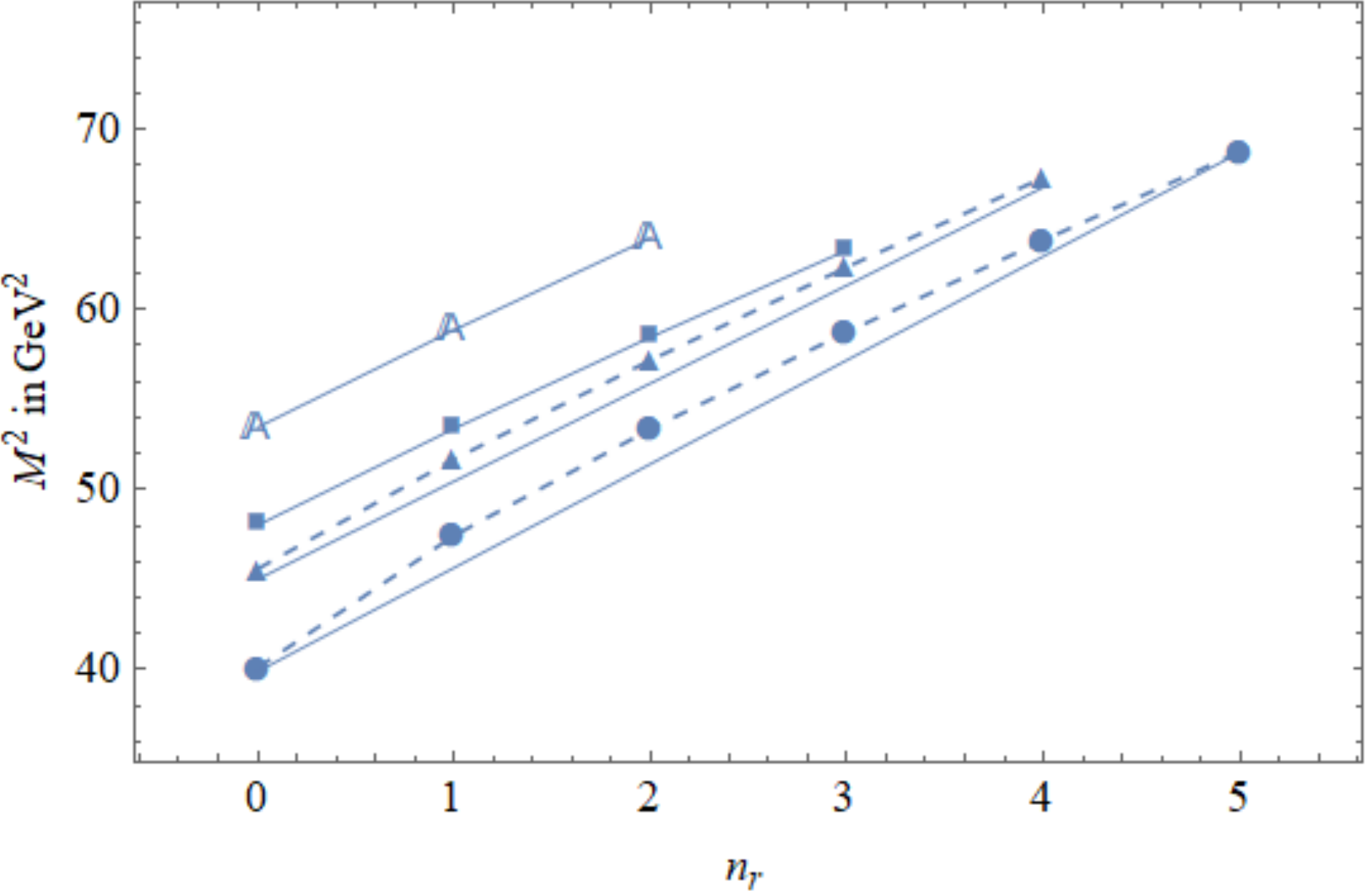}
\caption{Regge trajectory $(n_r,M^2)$ for the Pseudoscalar and vector S state, excited P and D state masses for $B_c$ meson.}
\label{fig:8}
\end{figure}

\begin{figure}
\centering
\includegraphics[width=0.5\textwidth]{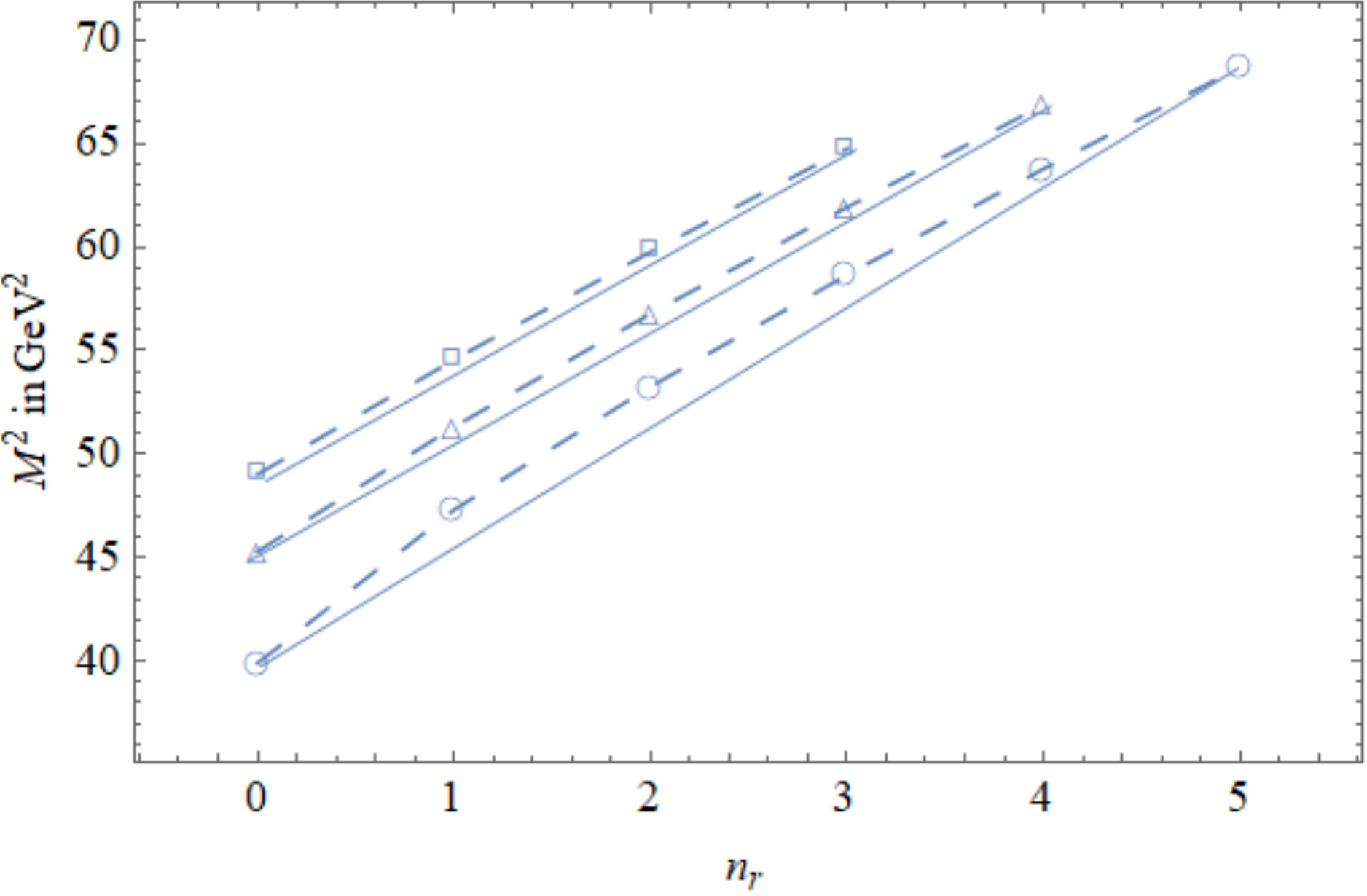}
\caption{Regge trajectory $(n_r,M^2)$  for the S-P-D States center of weight mass for $B_c$ meson.}
\label{fig:9}
\end{figure}

In Figures \ref{fig:7},\ref{fig:8},\ref{fig:9} we have plotted Regge trajectories for $B_c$ meson. The fitted slopes and intercepts for the trajectories can be found in Tables.\ref{Table:21},\ref{Table:22},$\&$,\ref{Table:23}. It is noted from our previous and present work that the parent and daughter nuclie Regge trajectories for $c\bar{c}$ meson are linear, but for $b\bar{b}$ and $B_c$ meson only the daughter nuclie trajectories are linear, but the parent nuclie trajectories are non-linear, this is because the $c\bar{c}$ and lighter meson size belong to the region where the confining term dominates in the inter-quark potential. But, the size of the first few excited states of $b\bar{b}$ and $B_c$ meson fall in the region where the coulomb part of the potential may play important part, as a result the parent nuclie Regge trajectories of $b\bar{b}$ and $B_c$ meson are non-linear, while the daughter trajectories are linear.

\begin{table*}
 \caption{S \& P wave mass spectrum(in GeV) of $B_c$ meson.}
 \begin{center}
 \begin{tabular}{ccccccccccc}
 \hline
State &without &with& \cite{Tang:2022xtx}& \cite{Wang:2022cxy} & \cite{Soni:2017wvy} &\cite{Devlani:2014nda}& \cite{Ebert:2011jc}& \cite{Godfrey:2004ya} & \cite{Davies:1996gi}&  Expt.\cite{ParticleDataGroup:2020ssz} \\
&correction&correction&Screening&Salpeter&NR&NR&RQM&RQM&Lattice&\\
\hline
$1^1S_0$ &6.456 &6.274&  6.262&6.277 & 6.272 & 6.278 & 6.272 & 6.271 &6.274$\pm$0.032  & 6.277$\pm$0.003\\
$1^3S_1$ &6.526 &6.332& 6.324&6.332& 6.321 & 6.331 & 6.333 & 6.338  &6.321$\pm$0.003& --\\
\hline
$2^1S_0$ &7.074 &6.851&  6.858&6.867 & 6.864 & 6.863 & 6.842 & 6.855 &6960$\pm$0.008& 6.841$\pm$0.006\cite{CMS:2019uhm} \\
$2^3S_1$ &7.102 &6.888& 6.881& 6.911 & 6.900 & 6.873 & 6.882 & 6.887 & 6990$\pm$0.008&6.871$\pm$0.001   \\
\hline
$3^1S_0$ &7.510 &7.275& 7.216&7.228&7.306 & 7.244 & 7.226  & --&--		 \\
$3^3S_1$ &7.534 &7.306& 7.232&7.272&7.338 & 7.249 & 7.258 & --	& --	 \\
\hline
$4^1S_0$ &7.884 &7.639& 7.500&--&7.684 & 7.564 & 7.585 		 & --&--		 \\
$4^3S_1$ &7.906 &7.666& 7.513&--&7.714 & 7.568 & 7.609 		 & --&--		 \\
\hline
$5^1S_0$ &8.220 &7.967& 7.743&--&8.025 & 7.852 & 7.928 		 & --	&--	 \\
$5^3S_1$ &8.240 &7.992& 7.754&--&8.054 & 7.855 & 7.947 		 & --	&--	 \\
\hline
$6^1S_0$ &8.530 &8.271& 7.959&--&8.340 & 8.120 & --				 & --& --    \\
$6^3S_1$ &8.550 &8.294& 7.968&-- &8.368 & 8.122 & --		 		 & -- &--  \\
\hline
$1^3P_0$ &6.870 &6.686& 6.883&6.705 &6.686 & 6.748 & 6.699 & 6.706  &6.727$\pm$0.003 &--\\
$1^1P_1$ &6.893 &6.709& 6.760&6.748&6.705 & 6.767 & 6.750 & 6.741  & 6.743$\pm$0.003&--\\
$1^3P_1$ &6.914 &6.731& 6.745&6.739&6.706 & 6.769 & 6.743 & 6.750  &6.743$\pm$0.003 &--\\
$1^3P_2$ &6.936 &6.752& 6.774&6.762&6.712 & 6.775 & 6.761 & 6.768  &6.765$\pm$0.003 &--\\
\hline
$2^3P_0$ &7.319 &7.113& 7.068&7.112 &7.146 & 7.139 & 7.094 & 7.122 & --&--\\
$2^1P_1$ &7.345 &7.139& 7.129&7.149&7.165 & 7.155 & 7.134 & 7.145  & --&--\\
$2^3P_1$ &7.370 &7.165& 7.116&7.144 &7.168 & 7.156 & 7.094 & 7.150  & --&--\\
$2^3P_2$ &7.395 &7.190& 7.141&7.163 &7.173 & 7.162 & 7.157 & 7.164  & --&--\\
\hline
$3^3P_0$ &7.701 &7.481& 7.432&7.408 &7.536 & 7.463 & 7.474 & --    			 & --&--\\
$3^1P_1$ &7.728 &7.508& 7.421&7.442 &7.555 & 7.479 & 7.510 & --    	 		 & --&--\\
$3^3P_1$ &7.755 &7.536& 7.408&7.440 &7.559 & 7.479 & 7.500 & --    			 & --&--\\
$3^3P_2$ &7.782 &7.563& 7.368&7.456 &7.565 & 7.485 & 7.524 & --    			 & --&--\\
\hline
$4^3P_0$ &8.042 &7.812& --&-- &7.885 & -- 		 & 7.817 & --		 		& --&--\\
$4^1P_1$ &8.071 &7.841& --&-- &7.905 & -- 		 & 7.853 & --				& --&--\\
$4^3P_1$ &8.100 &7.869& --&-- &7.908 & -- 		 & 7.844 & --				& --&-- \\
$4^3P_2$ &8.129 &7.898& --&--&7.915 & -- 		 & 7.867 & --		 		& --&--\\
\hline
 \end{tabular}
 \end{center}
 \label{Table:mass s,p}
 \end{table*}

 \begin{table*}
\caption{D \& F wave mass spectrum(in GeV) of $B_c$ meson.}
\begin{center}
\begin{tabular}{ccccccccc}
\hline
State  &without &with &\cite{Tang:2022xtx} &\cite{Wang:2022cxy}&\cite{Soni:2017wvy} &\cite{Devlani:2014nda}& \cite{Ebert:2011jc}& \cite{Godfrey:2004ya}   \\
&correction&correction&Screening&Salpeter&NR&NR&RQM&RQM\\
\hline
$1^3D_3$ &7.127 &7.050& 7.015&7.035&6.990 & 7.026 & 7.029 & 7.045	 \\
$1^1D_2$ &7.200 &7.003&  7.016&7.025 &6.994 & 7.035 & 7.026 & 7.041 	 \\
$1^3D_2$ &7.177 &6.979& 7.010&7.029 &6.997 & 7.025 & 7.025 & 7.036 	 \\
$1^3D_1$ &7.129 &6.930& 7.006&7.014&6.998 & 7.030 & 7.021 & 7.028	 \\
\hline
$2^3D_3$ &7.654 &7.441& &7.355 &7.399 & 7.363 & 7.405 & --   		\\
$2^1D_2$ &7.601 &7.387& &7.345&7.401 & 7.370 & 7.400 & --   		 \\
$2^3D_2$ &7.575 &7.361& &7.349 &7.403 & 7.361 & 7,399 & --   		 \\
$2^3D_1$ &7.521 &7.307& &7.335 &7.403 & 7.365 & 7.392 & --    		 \\
\hline
$3^3D_3$ &8.012 &7.787&-- &-- &7.761 & --    	& 7.750 &   --   		    \\
$3^1D_2$ &7.957 &7.731&-- & --&7.762 & --    	& 7.743 &   --   		    \\
$3^3D_2$ &7.929 &7.703& --&-- &7.764 & --    	& 7.741 &   --   		    \\
$3^3D_1$ &7.872 &7.647&-- &-- &7.762 & --    	& 7.732 &   --   		  \\
\hline
$4^3D_3$ &8.340 &8.107& --&--&8.092	& --		& --		& --			\\
$4^1D_2$ &8.282 &8.047&-- &-- &8.093 & --		& --		& --			\\
$4^3D_2$ &8.253 &8.017&-- &-- &8.094 & --		& --		& --				\\
$4^3D_1$ &8.194 &7.957& --&--&8.091 & --		& --		& --				\\
\hline
$1^3F_2$ &7.105 &7.131& 7.215&-- &7.234 & --    	& 7.273 & 7.269    	   \\
$1^3F_3$ &7.177 &7.209& 7.207&--&7.242 & --    	& 7.269 & 7.276    	    \\
$1^1F_3$ &7.200 &7.234& 7.219&-- &7.241 & --    	& 7.268 & 7.266    	   \\
$1^3F_4$ &7.270 &7.312& 7.210&-- &7.244 & --    	& 7.277 & 7.271	     \\
\hline
$2^3F_2$ &7.494 &7.477& --&-- &7.607 & --    	& 7.618 & --    		    \\
$2^3F_3$ &7.575 &7.561& --&-- &7.615 & --    	& 7.616 & --    		    \\
$2^1F_3$ &7.601 &7.589& --&--&7.614 & --    	& 7.615	& --    	    \\
$2^3F_4$ &7.681 &7.673&-- &--&7.617 & --    	& 7.617 & --    		  \\
\hline
\end{tabular}\end{center}
\label{Table:mass d,f}
\end{table*}


\begin{table}
    \caption{Fitted parameters of the $(J,M^2)$ Regge trajectory with natural and un-natural parity for $B_c$ meson}
    \begin{center}
    \begin{tabular}{cccc}
    \hline
    $\mathit b\bar{\mathit c}$& $\alpha (\mathit {GeV}^{-2})$ & $\alpha_0$ & \\
    \hline
    Parent& 0.169$\pm$0.007&-5.902$\pm$0.390&\\
    First Daughter&0.184$\pm$0.004&-7.467$\pm$0.250&\\
    Second Daughter&0.187$\pm$0.003&-8.342$\pm$0.190&\\
    Third Daughter&0.190$\pm$0.003&-9.172$\pm$0.185&\\
    \hline
    Parent& 0.169$\pm$0.007&-5.902$\pm$0.390&\\
    First Daughter&0.184$\pm$0.004&-7.467$\pm$0.250&\\
    Second Daughter&0.187$\pm$0.003&-8.342$\pm$0.190&\\
    Third Daughter&0.190$\pm$0.003&-9.172$\pm$0.185&\\
    \hline
    \end{tabular}\end{center}
    $\label{Table:21}$
    \end{table}

\begin{table}
    \caption{Fitted parameters of Regge trajectory $(n_r,M^2)$ for the Pseudoscalar and vector $S$ state, excited $P$ and $D$ state masses for $B_c$ meson}
    \begin{center}
    \begin{tabular}{cccc}
    \hline
    $\mathit b\bar{\mathit c}$& $\beta (\mathit {GeV}^{-2})$ & $\beta_0$ & \\
    \hline
    Parent& 0.176$\pm$0.006&-7.229$\pm$0.369&\\
    First Daughter&0.184$\pm$0.004&-8.468$\pm$0.250&\\
    Second Daughter&0.196$\pm$0.003&-9.446$\pm$0.181&\\
    Third Daughter&0.190$\pm$0.003&-10.172$\pm$0.185&\\
    \hline
    \end{tabular}\end{center}
    $\label{Table:22}$
    \end{table}

\begin{table}
\caption{Fitted parameters of Regge trajectory $(n_r,M^2)$ for the $S-P-D$ states center of weight mass for $B_c$ meson}
\begin{center}
\begin{tabular}{ccccc}
\hline
$b\bar c$ & $\beta(GeV^{-2})$ & $\beta_0$ &\\
\hline
S&0.175$\pm$0.007&-7.183$\pm$0.373&\\
P&0.180$\pm$0.004&-8.220$\pm$0.005&\\
D&0.191$\pm$0.003&-9.382$\pm$0.184&\\

\hline
\end{tabular}\end{center}
$\label{Table:23}$
\end{table}

\subsection{Decay Properties}
The decay constant of pseudoscalar and vector states considering the QCD correction, weak decays, radiative and pure leptonic decay width, and electromagnetic transition width of $B_c$ meson have been calculated using the computed masses, potential parameters, and normalize reduced wave-function. \\
The computed values of decay constant for different pseudoscalar and vector states can be found in Table.\ref{Table:decay constant pseudo scalar}$\&$\ref{Table:decay constant vector}. For, pseudoscalar states we compare our results with non-relativistic potential model results(NR)\cite{Rai:2006dt,Soni:2017wvy,Patel:2008na}, results from QCD sum rules\cite{Akbar:2020lof}. For, Vector states we compare with available non-relativistic potential model result only. Our calculated values are nearby the available theoretical values, but are little bit on the higher side.\\
Using the computed decay constant we estimate the weak decay widths in the spectator quark approximation and also calculate the lifetime of pseudoscalar $B_c$ meson in $\bar{c}s$ channel. For the present work the values of $\Gamma(b \rightarrow X)$ and $\Gamma(c \rightarrow X)$ are $6.7 \times 10^{-4} eV$ and $4.2 \times 10^{-4} eV$, respectively. $\Gamma(Anni.)$ for both $\Gamma(b \rightarrow X)$ and $\Gamma(c \rightarrow X)$ channels have been determined and used to calculate the lifetime($\tau$), the total decay width $\Gamma(B_C \rightarrow X)$, and branching ratio. The computed values of the lifetime and branching ratios can be found in Table\ref{Table:lifetime and branching ratio1}. The computed results are in accordance with available theoretical values.\\
In Table\ref{Table:pure leptonic decay width},\ref{Table:radiative leptonic decay width},$\&$ \ref{Table:br for pure and radiative leptonic} computed results of pure and radiative leptonic decay widths, and the corresponding branching ratios for pseudoscalar states are given. Our calculated values are of the same order when compared with other theoretical approaches. No exact conclusion can be reached due to limited studies about pure and radiative decay. More studies on these decays are needed.\\
We tabulate our results for electromagnetic transitions in Table\ref{Table:e1Bc}$\&$\ref{Table:m1Bc}, the calculated values are compatible with theoretical results.\\

\begin{table*}
 \caption{Pseudoscalar decay constants of the $B_c$ meson in MeV.}
 \begin{center}
 \begin{tabular}{cccccccc}
 \hline
state&$f_{corr}$&\cite{Rai:2006dt}&\cite{Soni:2017wvy}&\cite{Patel:2008na}&\cite{Veliev:2010vd}&\cite{Davies:1996gi}&\cite{Rai:2008sc}\\
&this work&NR&NR&QCD Sum Rule&Lattice&NR\\
\hline
1S&564.064&556&432.955&465&476$\pm$27&440$\pm$20&525\\
2S&451.592&--&355.504&361&--&\\
3S&410.885&--&325.659&319&--&\\
4S&386.218&--&307.492&293&--&\\
5S&368.653&--&294.434&275&--&\\
6S&354.874&--&284.237&261&--&\\
\hline
 \end{tabular} \end{center}
 \label{Table:decay constant pseudo scalar}
 \end{table*}

\begin{table*}
 \caption{Vector decay constants of the $B_c$ meson in MeV.}
 \begin{center}
 \begin{tabular}{cccccc}
 \hline
state&$f_{corr}$&\cite{Soni:2017wvy}&\cite{Patel:2008na}&\cite{Akbar:2020lof}&\cite{Rai:2008sc}\\
\hline
1S&531.905&434.642&435&484&528\\
2S&424.474&356.435&337&347&\\
3S&385.475&326.374&297&306&\\
4S&362.269&308.094&273&284&\\
5S&345.632&294.962&256&269&\\
6S&332.529&284.709&243&258&\\
\hline
 \end{tabular} \end{center}
 \label{Table:decay constant vector}
 \end{table*}

\begin{table*}
 \caption{Life-time and branching ratios(BR) of pseudo scalar  $B_c$ meson for $\bar{c}s$ channel.}
 \begin{center}
 \begin{tabular}{ccccc}
 \hline
state&Life-time & BR& BR&BR\\
& $10^{-12}$ sec.&$(b \rightarrow X)$&$(c \rightarrow X)$&$(Anni)$\\
\hline
1S(our result)&0.438&62&26&12.5\\
1S(Expt.)&$0.507\pm0.009$&\\
1S\cite{Soni:2017wvy}&$0.539$&\\
1S\cite{https://doi.org/10.48550/arxiv.2003.08491}&0.442&58&36.9&4.9\\
1S\cite{Godfrey:2004ya}&0.75&54&38&8\\
1S\cite{AbdEl-Hady:1998uiq}&0.47\\
1S\cite{Rai:2006dt}&0.47\\
\hline
 \end{tabular} \end{center}
 \label{Table:lifetime and branching ratio1}
 \end{table*}


\begin{table*}
\caption{Pure leptonic decay width of pseudo scalar $B_c$ meson($10^{-21} GeV$).}
\begin{center}
\begin{tabular}{cccccc}
\hline
lepton&state&present&\cite{Shah:2022pda}&\cite{Chang:1999gn}\\
&&work&\\
\hline
    &1S&5.007&2.829&1.827  \\
    &2S&3.504&--&--  \\
$e$ &3S&3.081&--&-- \\
    &4S&2.858&--&-- \\
    &5S&2.716&--&-- \\
\hline
     &1S&2.178&1.209&0.7841 \\
     &2S&1.524&--&-- \\
$\mu$&3S&1.304&--&--  \\
     &4S&1.243&--&--  \\
     &5S&1.181&--&--  \\
\hline
      &1S&5.122&2.732&1.862  \\
      &2S&3.687&--&--  \\
$\tau$&3S&3.294&--&--  \\
      &4S&3.092&--&--  \\
      &5S&2.965&--&--  \\
\hline
\end{tabular} \end{center}
\label{Table:pure leptonic decay width}
\end{table*}

\begin{table*}
\caption{Radiative leptonic decay width of pseudo scalar $B_c$ meson($10^{-16} GeV$).}
\begin{center}
\begin{tabular}{ccccc}
\hline
state&present&\cite{Shah:2022pda}\\
&work&\\
\hline
1S&2.534&7.851  \\
2S&2.665&-- \\
3S&3.088&--\\
4S&3.583&-- \\
5S&4.124&-- \\
\hline
\end{tabular} \end{center}
\label{Table:radiative leptonic decay width}
\end{table*}

\begin{table*}
\caption{Branching Ratio(BR) for Pure and Radiative leptonic decay for $1S$ pseudo scalar $B_c$ meson.}
\begin{center}
\begin{tabular}{ccccc}
\hline
&BR&BR&BR&BR\\
&$e$&$\mu$&$\tau$&radiative\\
\hline
Our result&$9.817 \times 10^{-9}$ & $4.270 \times 10^{-4}$ & $10.043 \times 10^{-2}$ & $4.969 \times 10^{-4}$\\
\cite{Shah:2022pda}& $2.192\times 10^{-9}$ & $0.937\times 10^{-4}$ & $2.117\times 10^{-2}$ & $6.086\times 10^{-5}$ \\
\cite{Chang:1999gn}& $1.44 \times 10^{-9}$ & $0.62\times 10^{-4}$ & $1.47\times 10^{-2}$ & $4.9\times 10^{-5}$ \\
\hline
\end{tabular} \end{center}
\label{Table:br for pure and radiative leptonic}
Note: - for calculating the Branching ratio we considered $B_c^+$ lifetime as $0.510 \times 10^{-12}$ seconds as per PDG\cite{ParticleDataGroup:2020ssz}.
\end{table*}

\begin{table*}
\caption{$E1$ transition widths of $B_c$ meson (in keV).}
\begin{center}\begin{tabular}{cccccc}
\hline
Transition & Present work &\cite{Soni:2017wvy}& \cite{Ebert:2003rh} & \cite{Godfrey:1985xj} & \cite{Devlani:2014nda}\\
\hline
$2^3S_1 \to 1^3P_0$ &1.289& 4.782	& 5.53	& 2.9		& 0.94	\\
$2^3S_1 \to 1^3P_1$ &9.70& 11.156	& 7.65	& 4.7		& 1.45	\\
$2^3S_1 \to 1^3P_2$ &4.608& 16.823	& 7.59	& 5.7		& 2.28	\\
$2^1S_0 \to 1^1P_1$ &6.920& 18.663	& 4.40	& 6.1		& 3.03	\\
\hline
$3^3S_1 \to 2^3P_0$ &22.760& 7.406	& --		& --		& --		\\
$3^3S_1 \to 2^3P_1$ &15.167& 17.049	& --		& --		& --		 \\
$3^3S_1 \to 2^3P_2$ &5.727& 25.112	& --		& --		& --		 \\
$3^1S_0 \to 2^1P_1$ &3.598& 27.988	& --		& --		& --		 \\
\hline
$1^3P_2 \to 1^3S_1$ &111.256& 55.761	& 122	& 83		& 64.24	\\
$1^3P_1 \to 1^3S_1$ &49.231& 53.294	& 87.1	& 11		& 51.14	\\
$1^3P_0 \to 1^3S_1$ &14.435& 46.862	& 75.5	& 55		& 58.55	\\
$1^1P_1 \to 1^1S_0$ &42.891& 71.923	& 18.4	& 80		& 72.28	\\
\hline
$2^3P_2 \to 2^3S_1$ &80.625& 41.259	& 75.3	& 55		& 64.92	 \\
$2^3P_1 \to 2^3S_1$ &47.473& 38.533	& 45.3	& 45		& 50.40	 \\
$2^3P_0 \to 2^3S_1$ &34.590& 38.308	& 34.0	& 42		& 55.05	 \\
$2^1P_1 \to 2^1S_0$ &63.020& 52.205	& 13.8	& 52		& 56.28	 \\
\hline
$2^3P_2 \to 1^3S_1$ &35.033& 60.195	& --		& 14		& --		 \\
$2^3P_1 \to 1^3S_1$ &39.985& 57.839	& --		& 5.4		& --		 \\
$2^3P_0 \to 1^3S_1$ &41.790& 52.508	& --		& 1.0		& --		 \\
$2^1P_1 \to 1^1S_0$ &37.683& 74.211	& --		& 19		& --		 \\
\hline
$1^3D_1 \to 1^3P_0$ &4.288& 44.783	& 133	& 55		& --		\\
$1^3D_1 \to 1^3P_1$ &3.403& 28.731	& 65.3	& 28		& --		\\
$1^3D_1 \to 1^3P_2$ &1.976& 1.786	& 3.82	& 1.8		& --		\\
$1^3D_2 \to 1^3P_1$ &5.433& 51.272	& 139	& 64		& --		 \\
$1^3D_2 \to 1^3P_2$ &3.629& 16.073	& 23.6	& 15		& --		 \\
$1^3D_3 \to 1^3P_2$ &6.514& 60.336 	& 149	& 78		& --		 \\
$1^1D_2 \to 1^1P_1$ &7.246& 66.020	& 143	& 63		& --		
\\ \hline \end{tabular} \end{center}
\label{Table:e1Bc}
\end{table*}

\begin{table*}
\caption{$M1$ transition widths of $B_c$ meson(in eV).}
\begin{center}\begin{tabular}{cccccc}
\hline
Transition & Present work &\cite{Soni:2017wvy}& \cite{Ebert:2003rh} & \cite{Godfrey:1985xj} &\cite{Devlani:2014nda} \\
\hline
$1^3S_1 \to 1^1S_0$ &42.021& 53.109 		& 33		& 80 		& 2.2		\\
$2^3S_1 \to 2^1S_0$ &15.052& 21.119 		& 17		& 10 		& 0.014	\\
$2^3S_1 \to 1^1S_0$ &475.78& 481.572 	& 428	& 600 	& 495	\\
$2^1S_0 \to 1^3S_1$ &490& 568.346		& 488	& 300 	& 1092	
\\ \hline \end{tabular} \end{center}
\label{Table:m1Bc}
\end{table*}

\section{Conclusion}
\label{sec:Conclusion}
In this article, we have calculated $B_c$ meson’s  mass spectra by taking into consideration relativistic correction in pNRQCD’s framework.  The model employed in this article suppresses the calculated masses of all states when compared to the masses calculated considering only the cornell potential. The calculated masses from present work match well with experimental and theoretically available masses. \\
Using the potential parameters, masses and normalised reduced wave function various decay properties like decay constant, weak decays, branching fraction, lifetime, pure and radiative leptonic decay widths, and electromagnetic transitions have been calculated. The computed decays appear to consent with available theoretical results from nor-relativistic potential models, QCD sum rules, and lattice QCD. Thus, justifying our choice of the potential for $B_c$ meson analysis.\\

\bibliographystyle{spphys}
\bibliography{EPJA}

\begin{thebibliography}{10}
\providecommand{\url}[1]{{#1}}
\providecommand{\urlprefix}{URL }
\expandafter\ifx\csname urlstyle\endcsname\relax
  \providecommand{\doi}[1]{DOI \discretionary{}{}{}#1}\else
  \providecommand{\doi}{DOI \discretionary{}{}{}\begingroup
  \urlstyle{rm}\Url}\fi

\bibitem{CDF:1998ihx}
F.~Abe, et~al., Phys. Rev. Lett. \textbf{81}, 2432 (1998).
\newblock \doi{10.1103/PhysRevLett.81.2432}

\bibitem{CDF:2005yjh}
A.~Abulencia, et~al., Phys. Rev. Lett. \textbf{96}, 082002 (2006).
\newblock \doi{10.1103/PhysRevLett.96.082002}

\bibitem{CDF:2007umr}
T.~Aaltonen, et~al., Phys. Rev. Lett. \textbf{100}, 182002 (2008).
\newblock \doi{10.1103/PhysRevLett.100.182002}

\bibitem{D0:2008bqs}
V.M. Abazov, et~al., Phys. Rev. Lett. \textbf{101}, 012001 (2008).
\newblock \doi{10.1103/PhysRevLett.101.012001}

\bibitem{LHCb:2012ihf}
R.~Aaij, et~al., Phys. Rev. Lett. \textbf{109}, 232001 (2012).
\newblock \doi{10.1103/PhysRevLett.109.232001}

\bibitem{ATLAS:2014lga}
G.~Aad, et~al., Phys. Rev. Lett. \textbf{113}(21), 212004 (2014).
\newblock \doi{10.1103/PhysRevLett.113.212004}

\bibitem{CMS:2019uhm}
A.M. Sirunyan, et~al., Phys. Rev. Lett. \textbf{122}(13), 132001 (2019).
\newblock \doi{10.1103/PhysRevLett.122.132001}

\bibitem{LHCb:2019bem}
R.~Aaij, et~al., Phys. Rev. Lett. \textbf{122}(23), 232001 (2019).
\newblock \doi{10.1103/PhysRevLett.122.232001}

\bibitem{CMS:2020rcj}
A.M. Sirunyan, et~al., Phys. Rev. D \textbf{102}(9), 092007 (2020).
\newblock \doi{10.1103/PhysRevD.102.092007}

\bibitem{Martin-Gonzalez:2022qwd}
B.~Mart\'\i{}n-Gonz\'alez, P.G. Ortega, D.R. Entem, F.~Fern\'andez, J.~Segovia,
    (2022)

\bibitem{Wang:2022cxy}
G.L. Wang, T.~Wang, Q.~Li, C.H. Chang, JHEP \textbf{05}, 006 (2022).
\newblock \doi{10.1007/JHEP05(2022)006}

\bibitem{Tang:2022xtx}
L.~Tang, T.y. Li, C.h. Wang, C.q. Pang, X.~Liu,   (2022)

\bibitem{Mansour:2021rru}
H.~Mansour, A.~Gamal, Results Phys. \textbf{33}, 105203 (2022).
\newblock \doi{10.1016/j.rinp.2022.105203}

\bibitem{Chang:1992bb}
C.H. Chang, Y.Q. Chen, Phys. Rev. D \textbf{46}, 3845 (1992).
\newblock \doi{10.1103/PhysRevD.46.3845}.
\newblock [Erratum: Phys.Rev.D 50, 6013 (1994)]

\bibitem{Chang:1991bp}
C.H. Chang, Y.Q. Chen, Phys. Lett. B \textbf{284}, 127 (1992).
\newblock \doi{10.1016/0370-2693(92)91937-5}

\bibitem{Braaten:1993jn}
E.~Braaten, K.m. Cheung, T.C. Yuan, Phys. Rev. D \textbf{48}(11), R5049 (1993).
\newblock \doi{10.1103/PhysRevD.48.R5049}

\bibitem{Cheung:1993qi}
K.m. Cheung, Phys. Rev. Lett. \textbf{71}, 3413 (1993).
\newblock \doi{10.1103/PhysRevLett.71.3413}

\bibitem{Chang:1992jb}
C.H. Chang, Y.Q. Chen, Phys. Rev. D \textbf{48}, 4086 (1993).
\newblock \doi{10.1103/PhysRevD.48.4086}

\bibitem{Chang:2001pm}
C.H. Chang, Y.Q. Chen, G.L. Wang, H.S. Zong, Phys. Rev. D \textbf{65}, 014017
  (2002).
\newblock \doi{10.1103/PhysRevD.65.014017}

\bibitem{Colangelo:1999zn}
P.~Colangelo, F.~De~Fazio, Phys. Rev. D \textbf{61}, 034012 (2000).
\newblock \doi{10.1103/PhysRevD.61.034012}

\bibitem{Qiao:2012hp}
C.F. Qiao, P.~Sun, D.~Yang, R.L. Zhu, Phys. Rev. D \textbf{89}(3), 034008
  (2014).
\newblock \doi{10.1103/PhysRevD.89.034008}

\bibitem{Ivanov:2006ni}
M.A. Ivanov, J.G. Korner, P.~Santorelli, Phys. Rev. D \textbf{73}, 054024
  (2006).
\newblock \doi{10.1103/PhysRevD.73.054024}

\bibitem{Chang:1992pt}
C.H. Chang, Y.Q. Chen, Phys. Rev. D \textbf{49}, 3399 (1994).
\newblock \doi{10.1103/PhysRevD.49.3399}

\bibitem{Kiselev:1993ea}
V.V. Kiselev, A.V. Tkabladze, Phys. Rev. D \textbf{48}, 5208 (1993).
\newblock \doi{10.1103/PhysRevD.48.5208}

\bibitem{Liu:1997hr}
J.F. Liu, K.T. Chao, Phys. Rev. D \textbf{56}, 4133 (1997).
\newblock \doi{10.1103/PhysRevD.56.4133}

\bibitem{Kiselev:2000pp}
V.V. Kiselev, A.E. Kovalsky, A.K. Likhoded, Nucl. Phys. B \textbf{585}, 353
  (2000).
\newblock \doi{10.1016/S0550-3213(00)00386-2}

\bibitem{Gregory:2009hq}
E.B. Gregory, C.T.H. Davies, E.~Follana, E.~Gamiz, I.D. Kendall, G.P. Lepage,
  H.~Na, J.~Shigemitsu, K.Y. Wong, Phys. Rev. Lett. \textbf{104}, 022001
  (2010).
\newblock \doi{10.1103/PhysRevLett.104.022001}

\bibitem{Dowdall:2012ab}
R.J. Dowdall, C.T.H. Davies, T.C. Hammant, R.R. Horgan, Phys. Rev. D
  \textbf{86}, 094510 (2012).
\newblock \doi{10.1103/PhysRevD.86.094510}

\bibitem{Godfrey:1985xj}
S.~Godfrey, N.~Isgur, Phys. Rev. \textbf{D32}, 189 (1985).
\newblock \doi{10.1103/PhysRevD.32.189}

\bibitem{Chen:1992fq}
Y.Q. Chen, Y.P. Kuang, Phys. Rev. D \textbf{46}, 1165 (1992).
\newblock \doi{10.1103/PhysRevD.47.350}.
\newblock [Erratum: Phys.Rev.D 47, 350 (1993)]

\bibitem{Fulcher:1998ka}
L.P. Fulcher, Phys. Rev. D \textbf{60}, 074006 (1999).
\newblock \doi{10.1103/PhysRevD.60.074006}

\bibitem{Gershtein:1994dxw}
S.S. Gershtein, V.V. Kiselev, A.K. Likhoded, A.V. Tkabladze, Phys. Rev. D
  \textbf{51}, 3613 (1995).
\newblock \doi{10.1103/PhysRevD.51.3613}

\bibitem{PhysRevD.52.5229}
J.~Zeng, J.W. Van~Orden, W.~Roberts, Phys. Rev. D \textbf{52}, 5229 (1995)

\bibitem{Gupta:1995ps}
S.N. Gupta, J.M. Johnson, Phys. Rev. D \textbf{53}, 312 (1996).
\newblock \doi{10.1103/PhysRevD.53.312}

\bibitem{Ebert:2002pp}
D.~Ebert, R.N. Faustov, V.O. Galkin, Phys. Rev. \textbf{D67}, 014027 (2003).
\newblock \doi{10.1103/PhysRevD.67.014027}

\bibitem{Ikhdair:2003ry}
S.M. Ikhdair, R.~Sever, Int. J. Mod. Phys. A \textbf{19}, 1771 (2004).
\newblock \doi{10.1142/S0217751X0401780X}

\bibitem{Godfrey:2004ya}
S.~Godfrey, Phys. Rev. \textbf{D70}, 054017 (2004).
\newblock \doi{10.1103/PhysRevD.70.054017}

\bibitem{Ikhdair:2004hg}
S.M. Ikhdair, R.~Sever, Int. J. Mod. Phys. \textbf{A20}, 4035 (2005).
\newblock \doi{10.1142/S0217751X05022275}

\bibitem{ATLAS:2021moa}
G.~Aad, et~al., Eur. Phys. J. C \textbf{81}(12), 1118 (2021).
\newblock \doi{10.1140/epjc/s10052-021-09749-7}

\bibitem{Wang:2013cha}
Z.G. Wang, Eur. Phys. J. C \textbf{73}(9), 2559 (2013).
\newblock \doi{10.1140/epjc/s10052-013-2559-7}

\bibitem{Chang_2021}
L.~Chang, M.~Chen, X.~qian Li, Y.~xin Liu, K.~Raya, Few-Body Systems
  \textbf{62}(1) (2021).
\newblock \doi{10.1007/s00601-020-01586-w}.
\newblock \urlprefix\url{https://doi.org/10.1007%2Fs00601-020-01586-w}

\bibitem{Chen:2020ecu}
M.~Chen, L.~Chang, Y.x. Liu, Phys. Rev. D \textbf{101}(5), 056002 (2020).
\newblock \doi{10.1103/PhysRevD.101.056002}

\bibitem{Yin:2019bxe}
P.L. Yin, C.~Chen, G.a. Krein, C.D. Roberts, J.~Segovia, S.S. Xu, Phys. Rev. D
  \textbf{100}(3), 034008 (2019).
\newblock \doi{10.1103/PhysRevD.100.034008}

\bibitem{Brambilla:2000db}
N.~Brambilla, A.~Vairo, Phys. Rev. D \textbf{62}, 094019 (2000).
\newblock \doi{10.1103/PhysRevD.62.094019}

\bibitem{Penin:2004xi}
A.A. Penin, A.~Pineda, V.A. Smirnov, M.~Steinhauser, Phys. Lett. B
  \textbf{593}, 124 (2004).
\newblock \doi{10.1016/j.physletb.2004.04.066}.
\newblock [Erratum: Phys.Lett.B 677, 343 (2009)]

\bibitem{Peset_2018}
C.~Peset, A.~Pineda, J.~Segovia, Journal of High Energy Physics
  \textbf{2018}(9) (2018).
\newblock \doi{10.1007/jhep09(2018)167}.
\newblock \urlprefix\url{https://doi.org/10.1007%2Fjhep09%282018%29167}

\bibitem{Peset:2018jkf}
C.~Peset, A.~Pineda, J.~Segovia, Phys. Rev. D \textbf{98}(9), 094003 (2018).
\newblock \doi{10.1103/PhysRevD.98.094003}

\bibitem{Allison:2004be}
I.F. Allison, C.T.H. Davies, A.~Gray, A.S. Kronfeld, P.B. Mackenzie, J.N.
  Simone, Phys. Rev. Lett. \textbf{94}, 172001 (2005).
\newblock \doi{10.1103/PhysRevLett.94.172001}

\bibitem{Mathur:2018epb}
N.~Mathur, M.~Padmanath, S.~Mondal, Phys. Rev. Lett. \textbf{121}(20), 202002
  (2018).
\newblock \doi{10.1103/PhysRevLett.121.202002}

\bibitem{Sauli:2011aa}
V.~Sauli, Phys. Rev. D \textbf{86}, 096004 (2012).
\newblock \doi{10.1103/PhysRevD.86.096004}

\bibitem{Leitao:2014jha}
S.~Leit\~ao, A.~Stadler, M.T. Pe\~na, E.P. Biernat, Phys. Rev. D
  \textbf{90}(9), 096003 (2014).
\newblock \doi{10.1103/PhysRevD.90.096003}

\bibitem{Fischer:2014cfa}
C.S. Fischer, S.~Kubrak, R.~Williams, Eur. Phys. J. \textbf{A51}, 10 (2015).
\newblock \doi{10.1140/epja/i2015-15010-7}

\bibitem{Deng:2016stx}
W.J. Deng, H.~Liu, L.C. Gui, X.H. Zhong, Phys. Rev. \textbf{D95}(3), 034026
  (2017).
\newblock \doi{10.1103/PhysRevD.95.034026}

\bibitem{Deng:2016ktl}
W.J. Deng, H.~Liu, L.C. Gui, X.H. Zhong, Phys. Rev. \textbf{D95}(7), 074002
  (2017).
\newblock \doi{10.1103/PhysRevD.95.074002}

\bibitem{Segovia:2016xqb}
J.~Segovia, P.G. Ortega, D.R. Entem, F.~Fernández, Phys. Rev. \textbf{D93}(7),
  074027 (2016).
\newblock \doi{10.1103/PhysRevD.93.074027}

\bibitem{Soni:2017wvy}
N.R. Soni, B.R. Joshi, R.P. Shah, H.R. Chauhan, J.N. Pandya, Eur. Phys. J.
  \textbf{C78}(7), 592 (2018).
\newblock \doi{10.1140/epjc/s10052-018-6068-6}

\bibitem{Devlani:2014nda}
N.~Devlani, V.~Kher, A.K. Rai, Eur. Phys. J. \textbf{A50}(10), 154 (2014).
\newblock \doi{10.1140/epja/i2014-14154-2}

\bibitem{Brambilla:2010cs}
N.~Brambilla, et~al., Eur. Phys. J. \textbf{C71}, 1534 (2011).
\newblock \doi{10.1140/epjc/s10052-010-1534-9}

\bibitem{Brambilla:2014jmp}
N.~Brambilla, et~al., Eur. Phys. J. \textbf{C74}(10), 2981 (2014).
\newblock \doi{10.1140/epjc/s10052-014-2981-5}

\bibitem{Chaturvedi:2019usm}
R.~Chaturvedi, A.K. Rai, Int. J. Theor. Phys. \textbf{59}(11), 3508 (2020).
\newblock \doi{10.1007/s10773-020-04613-y}

\bibitem{Chaturvedi:2018tjr}
R.~Chaturvedi, N.R. Soni, J.N. Pandya, A.K. Rai, J. Phys. G \textbf{47}(11),
  115003 (2020).
\newblock \doi{10.1088/1361-6471/abaa99}

\bibitem{Isgur:1978xj}
N.~Isgur, G.~Karl, Phys. Rev. D \textbf{18}, 4187 (1978).
\newblock \doi{10.1103/PhysRevD.18.4187}

\bibitem{VijayaKumar:1997pg}
K.B. Vijaya~Kumar, A.K. Rath, S.B. Khadkikar, Pramana \textbf{48}, 997 (1997).
\newblock \doi{10.1007/BF02847459}

\bibitem{Gupta:1994mw}
S.N. Gupta, J.M. Johnson, Phys. Rev. \textbf{D51}, 168 (1995).
\newblock \doi{10.1103/PhysRevD.51.168}

\bibitem{article}
Y.~Koma, M.~Koma, Few-Body Systems \textbf{54} (2013).
\newblock \doi{10.1007/s00601-012-0542-8}

\bibitem{Patrignani:2016xqp}
C.~Patrignani, et~al., Chin. Phys. \textbf{C40}(10), 100001 (2016).
\newblock \doi{10.1088/1674-1137/40/10/100001}

\bibitem{Parmar:2010ii}
A.~Parmar, B.~Patel, P.C. Vinodkumar, Nucl. Phys. \textbf{A848}, 299 (2010).
\newblock \doi{10.1016/j.nuclphysa.2010.08.016}

\bibitem{Rai:2008sc}
A.K. Rai, B.~Patel, P.C. Vinodkumar, Phys. Rev. \textbf{C78}, 055202 (2008).
\newblock \doi{10.1103/PhysRevC.78.055202}

\bibitem{Patel:2008na}
B.~Patel, P.C. Vinodkumar, J. Phys. \textbf{G36}, 035003 (2009).
\newblock \doi{10.1088/0954-3899/36/3/035003}

\bibitem{VanRoyen:1967nq}
R.~Van~Royen, V.F. Weisskopf, Nuovo Cim. A \textbf{50}, 617 (1967).
\newblock \doi{10.1007/BF02823542}.
\newblock [Erratum: Nuovo Cim.A 51, 583 (1967)]

\bibitem{Braaten:1995ej}
E.~Braaten, S.~Fleming, Phys. Rev. \textbf{D52}, 181 (1995).
\newblock \doi{10.1103/PhysRevD.52.181}

\bibitem{Berezhnoy:1996an}
A.V. Berezhnoy, V.V. Kiselev, A.K. Likhoded, Z. Phys. A \textbf{356}, 89
  (1996).
\newblock \doi{10.1007/s002180050152}

\bibitem{AbdEl-Hady:1998uiq}
A.~Abd El-Hady, M.A.K. Lodhi, J.P. Vary, Phys. Rev. D \textbf{59}, 094001
  (1999).
\newblock \doi{10.1103/PhysRevD.59.094001}

\bibitem{Chang:1997re}
C.H. Chang, J.P. Cheng, C.D. Lu, Phys. Lett. B \textbf{425}, 166 (1998).
\newblock \doi{10.1016/S0370-2693(98)00177-4}

\bibitem{Radford:2009qi}
S.F. Radford, W.W. Repko, Nucl. Phys. \textbf{A865}, 69 (2011).
\newblock \doi{10.1016/j.nuclphysa.2011.06.032}

\bibitem{Eichten:1974af}
E.~Eichten, K.~Gottfried, T.~Kinoshita, J.B. Kogut, K.D. Lane, T.M. Yan, Phys.
  Rev. Lett. \textbf{34}, 369 (1975).
\newblock \doi{10.1103/PhysRevLett.34.369, 10.1103/PhysRevLett.36.1276}.
\newblock [Erratum: Phys. Rev. Lett.36,1276(1976)]

\bibitem{Eichten:1978tg}
E.~Eichten, K.~Gottfried, T.~Kinoshita, K.D. Lane, T.M. Yan, Phys. Rev.
  \textbf{D17}, 3090 (1978).
\newblock \doi{10.1103/PhysRevD.17.3090, 10.1103/physrevd.21.313.2}.
\newblock [Erratum: Phys. Rev.D21,313(1980)]

\bibitem{Pandya:2014qma}
J.N. Pandya, N.R. Soni, N.~Devlani, A.K. Rai, Chin. Phys. \textbf{C39}(12),
  123101 (2015).
\newblock \doi{10.1088/1674-1137/39/12/123101}

\bibitem{Davies:1996gi}
C.T.H. Davies, K.~Hornbostel, G.P. Lepage, A.J. Lidsey, J.~Shigemitsu, J.H.
  Sloan, Phys. Lett. B \textbf{382}, 131 (1996).
\newblock \doi{10.1016/0370-2693(96)00650-8}

\bibitem{Ebert:2011jc}
D.~Ebert, R.N. Faustov, V.O. Galkin, Eur. Phys. J. \textbf{C71}, 1825 (2011).
\newblock \doi{10.1140/epjc/s10052-011-1825-9}

\bibitem{ParticleDataGroup:2020ssz}
P.A. Zyla, et~al., PTEP \textbf{2020}(8), 083C01 (2020).
\newblock \doi{10.1093/ptep/ptaa104}

\bibitem{Rai:2006dt}
A.K. Rai, P.C. Vinodkumar, Pramana \textbf{66}, 953 (2006).
\newblock \doi{10.1007/BF02704795}

\bibitem{Akbar:2020lof}
N.~Akbar, Phys. Atom. Nucl. \textbf{83}(6), 899 (2020).
\newblock \doi{10.1134/S1063778820060034}

\bibitem{Veliev:2010vd}
E.V. Veliev, K.~Azizi, H.~Sundu, N.~Aksit, J. Phys. \textbf{G39}, 015002
  (2012).
\newblock \doi{10.1088/0954-3899/39/1/015002}

\bibitem{https://doi.org/10.48550/arxiv.2003.08491}
N.~Akbar.
\newblock Decay properties of conventional and hybrid $b_c$ mesons (2020).
\newblock \doi{10.48550/ARXIV.2003.08491}.
\newblock \urlprefix\url{https://arxiv.org/abs/2003.08491}

\bibitem{Shah:2022pda}
M.~Shah, R.~Patel, B.~Pandya, A.~Majethiya, P.C. Vinodkumar, in \emph{{65th DAE
  BRNS Symposium on nuclear physics}} (2022)

\bibitem{Chang:1999gn}
C.H. Chang, C.D. Lu, G.L. Wang, H.S. Zong, Phys. Rev. D \textbf{60}, 114013
  (1999).
\newblock \doi{10.1103/PhysRevD.60.114013}

\bibitem{Ebert:2003rh}
D.~Ebert, R.N. Faustov, V.O. Galkin, Mod. Phys. Lett. \textbf{A18}, 1597
  (2003).
\newblock \doi{10.1142/S0217732303011307}

\end{thebibliography}

\end{document}